\documentclass[aps,floatfix,superscriptaddress,notitlepage]{
revtex4}
\usepackage{amsmath,amssymb,amsfonts,graphics,graphicx,dcolumn,bm,enumerate}
\usepackage{comment,natbib,appendix}
\usepackage{multirow,color}

\usepackage{amsthm}
\usepackage{epsfig}
\usepackage{natbib}
\usepackage{hyperref}
\usepackage{graphicx}
\usepackage{epstopdf}

\newcommand{\cmp}
{\affiliation{Condensed Matter Physics Division, 
Saha Institute of Nuclear Physics, 1/AF Bidhannagar, Kolkata 700064, India.}}
\newcommand{\isi}
{\affiliation{Economic Research Unit, Indian Statistical Institute, 203 B. T. Road, Kolkata 700108, India.}}
\newcommand{\aalto}
{\affiliation{Department of Computer Science, Aalto University School of 
Science, P.O. Box 15400, FI-00076 AALTO, Finland}}

\begin{document}

\title{Socio-economic inequality and prospects of institutional Econophysics}

\author{Arnab Chatterjee}%
\email[Email: ]{arnabchat@gmail.com} 
\cmp
\author{Asim Ghosh}
\email[Email: ]{asim.ghosh@aalto.fi}
\aalto
\author{Bikas K Chakrabarti}%
\email[Email: ]{bikask.chakrabarti@saha.ac.in}
\cmp \isi

\begin{abstract}
Socio-economic inequality is measured  using various indices.
 The Gini ($g$) index, giving the overall inequality is the most commonly used, 
 while the recently introduced Kolkata ($k$) index gives a measure of $1-k$ 
 fraction of population who possess top  $k$ fraction of wealth in the society.
 This article reviews the character of such inequalities, as seen from a variety of data sources, 
 the apparent relationship between the two indices, and what toy models tell us. These socio-economic 
 inequalities are also investigated in the context of man-made social conflicts or wars,
  as well as in natural disasters. Finally, we forward a proposal for an international institution with 
  sufficient fund for visitors, where natural and social scientists from various 
  institutions of the world can come to discuss, debate and formulate further developments.
 \end{abstract}

 \maketitle
 
\section{Introduction: Socio-economic inequality}
The complex dynamics of human social interactions lead to interesting phenomena, 
and inequalities at various levels often show up in course.
The recent availability of huge amount of data (empirical data from databases, electronic footprints,
and sometimes large surveys) for various forms of human social interactions has made it easier
to uncover certain patterns present, to analyze them and investigate the reasons behind various 
socio-economic inequalities manifested in them.
The age of \textit{Big data} has opened up new avenues and challenges, and scientists are in the quest to
understand \textit{why} certain things look like as they do and \textit{how} do they happen.
Researchers are pooling in 
knowledge and techniques from various disciplines~\cite{lazer09}, e.g., 
statistics, applied mathematics, information theory, computer science, 
while tools of statistical physics have proved to be quite successful in 
better understanding of the precise (spatio-temporal) nature and  
origin of socio-economic inequalities prevalent in our society.
More the data that is acquired and analyzed, more we become confident in addressing the 
\textit{why}s, and \textit{how}s.

Statistical physics tells us that systems of a large number of interacting dynamical 
units collectively exhibit a behavior which is solely determined by only a few basic 
dynamical properties of the individual constituent units and of the embedding dimension,
but is independent of all other details.
This feature, which is specific to `critical phenomena', as in continuous phase 
transitions, is known as 
\textit{universality}~\cite{stanley1971introduction}. 
There is no shortage of empirical evidence that several social phenomena are 
characterized by simple emergent behavior out of the interactions of many individual constituent units. 
In recent  times, a growing community of researchers have been analyzing 
large-scale social dynamics to uncover universal patterns and proposing simple 
microscopic models to describe them, very similar to the minimalistic models used in statistical physics 
to understand physical phenomena.
These studies have revealed quite a few interesting patterns and behaviors in social 
systems, as in 
elections~\cite{fortunato2007scaling,chatterjee2013universality,
mantovani2011scaling}, 
population growth~\cite{rozenfeld2008laws} and 
economic growth~\cite{stanley1996scaling},
income and wealth distributions \cite{chakrabarti2013econophysics}, 
financial markets~\cite{mantegna2000introduction}, 
languages~\cite{petersen2012statistical}, etc.
(see Refs.~\cite{Castellano:2009,Sen:2013} for a review).

Socio-economic 
inequality~\cite{arrow2000meritocracy,stiglitz2012price,neckerman2004social,
goldthorpe2010analysing,chatterjee2015sociorev} 
usually concerns the existence of unequal `wealth' and `fortunes'
accumulated due to complex dynamics and interactions within the society. Usually 
containing structured and recurrent patterns of unequal distributions of goods, 
wealth, opportunities, and even rewards and punishments, this is classically 
measured in terms of  \textit{inequality of conditions},
and \textit{inequality of opportunities}.
The former refers to the unequal distribution of income,
wealth, assets and material goods. 
while the latter refers to the unequal distribution of `life
chances'. This is reflected in levels of education, health 
status, treatment done by the criminal justice system etc. Socio-economic 
inequalities are mostly responsible for conflicts, wars, crises, oppressions, criminal 
activities, instability in political scenario and socio-political unrest, and that in turn affects 
economic growth~\cite{hurst1995social}.
Historically as well as traditionally, economic inequalities have been extensively studied in the 
context of income and 
wealth~\cite{yakovenko2009colloquium,chakrabarti2013econophysics,
aoyama2010econophysics},
although it is also measured for many quantities like energy 
consumption~\cite{lawrence2013global}.
The studies of inequality in 
society~\cite{piketty2014inequality,Cho23052014,Chin23052014,Xie23052014} has 
been always very important, and is also a topic of contemporary focus and immediate 
global interest, drawing attention of researchers across disciplines -- 
economics, sociology, mathematics, statistics, demography, geography, graph 
theory, computer science, and not surprisingly, theoretical physics.

Quantifying socio-economic inequalities is a challenge, but is done in numerous ways. 
The probability distributions of various quantities, of course provide the most detailed measures.
It is very common to find that most quantities display broad distributions -- most common are log-normals, 
power-laws or their combinations.
For example, the distribution of income is usually found to be exponential for the bulk followed by a power 
law~\cite{druagulescu2001exponential,chakrabarti2013econophysics} for the top income range.
However, such distributions can widely differ in their forms and 
details, and as such they are rather difficult to handle. This leads to 
the introduction of various \textit{indices} like the Gini~\cite{gini1921measurement}, 
Theil~\cite{theil1967economics}, Pietra~\cite{eliazar2010measuring} and other 
socio-geometric 
indices~\cite{eliazar2015asociogeometry,eliazar2015bsociogeometry}, which try 
to characterize various geometric features of these distributions using a single number.
Of course, each of these indices come with certain merits, and certain indices are more useful than others, 
depending on the context they are used in.
In this article we will focus on the most common one, the Gini index and a recently 
proposed $k$ index ($k=$ Kolkata) which has a nice, useful socio-geometric 
interpretation.

\begin{figure}[t]
\centering
\includegraphics[width=10.0cm]{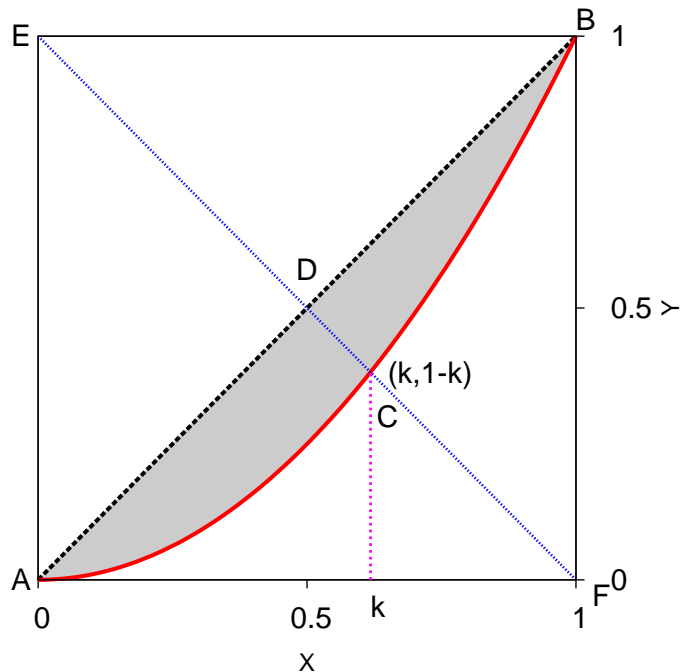}
 \caption{Lorenz curve is shown in solid red line for a typical probability 
distribution function and the equality line in dotted black diagonal.
The Lorenz curve shows the cumulative fraction of `wealth' possessed by the 
corresponding fraction of poorer population.
The $g$-index is given by area of the shaded region, while the 
$k$-index is computed from the coordinate of the point of intersection C 
($k,1-k$) of the Lorenz curve and the diagonal perpendicular to 
the equality line.
Thus, while the $g$-index measures the overall inequality in the system, the 
$k$-index gives the fraction $k$ of wealth possessed by the $1-k$ fraction 
of richer population.
 }
 \label{fig:lorenz}
\end{figure}

The most commonly used measure  to quantify socio-economic inequality is the Gini index. 
To compute this, one has to consider the `Lorenz curve'~\cite{Lorenz}, which shows 
the cumulative proportion $X$ of (poor to rich) ordered individuals 
(entries) in terms of the cumulative share $Y$ of their wealth.
$Y$ can of course represent income or assets of individuals but it can as well 
represent citation of articles, votes in favor of candidates, population of 
cities etc.
It is first computed from a given statistical distribution or a dataset.
The Gini index ($g$), defined as the ratio of the area enclosed between 
the Lorenz curve and the equality line, to that below the equality line,
taking values $0$ for absolute equality and $1$ for absolute inequality. 
Let the area between 
(i) the Lorenz curve and the equality line be represented as $\cal{A}$, and 
(ii) that below the Lorenz curve be $\cal{B}$  (See Fig.~\ref{fig:lorenz}). Then
the Gini index is $g=\cal{A}/(\cal{A+B})=$ $2\cal{A}$.
The recently introduced Kolkata index (symbolizing 
the extreme nature of social inequalities in Kolkata) or `$k$-index'~\cite{ghosh2014inequality}, 
is defined as the fraction $k$ such that  $(1-k)$ fraction 
of people (or papers) possess $k$ fraction of highest incomes 
(or citations)~\cite{inoue2015measuring,chatterjee2016universality,ghosh2016inequality}.

\begin{figure}[h]
\includegraphics[width=12.5cm]{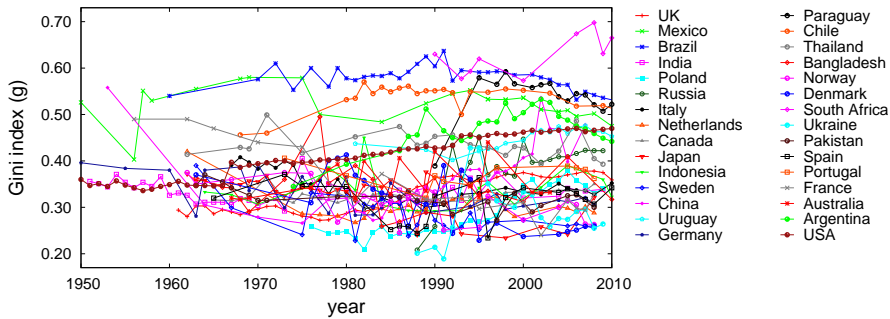}
\caption{Gini index from World Bank data~\cite{worldbank} for income for several 
countries over years.
}
\label{fig:emp_gini}
\end{figure}
The empirical data on Gini index from World Bank data~\cite{worldbank} for incomes 
over several years are given in FIg.~\ref{fig:emp_gini}. The values seem to be 
mostly between $0.2$ and $0.6$. In the later part of our article we will argue that 
the simple kinetic exchange models can even reproduce this feature.

We also discuss here the specific case of the citation distributions.
It was shown earlier~\cite{Radicchi:2008} that the 
distribution of citations $c$
to papers within a discipline has a broad distribution, which is universal across 
broad scientific
disciplines, by defining a relative indicator $c_f=c/\langle c \rangle$, where 
$\langle c \rangle$ is the 
average citation within a discipline.
Our study~\cite{chatterjee2016universality} confirmed this case for academic 
institutions as well as journals across disciplines.

Studies on the statistics of human deaths from wars, conflicts and 
natural disasters shows that the form of the probability distribution for number of people killed 
exhibit power law decay for the largest sizes,  the exponent
values being quite similar. We argue if a common mechanism is responsible for similarity that is manifested.

\section{Introduction: Institutional econophysics and sociophysics}

In view of the truly interdisciplinary nature of econophysics and sociophysics, it can
be argued that some interdisciplinary visiting facilities for social and natural scientists are absolutely 
necessary today. These will provide scientists from different disciplines to interact over some long period, 
discuss and debate and develop in their own discipline.
In the concluding part of this article, we argue about the need to establish a 
research institute dedicated to socio-economic problems with an interdisciplinary 
character, with some specific model in mind.

\section{Inequality in citations for academic institutions and journals}
\begin{figure}[h]
\includegraphics[width=6.0cm]{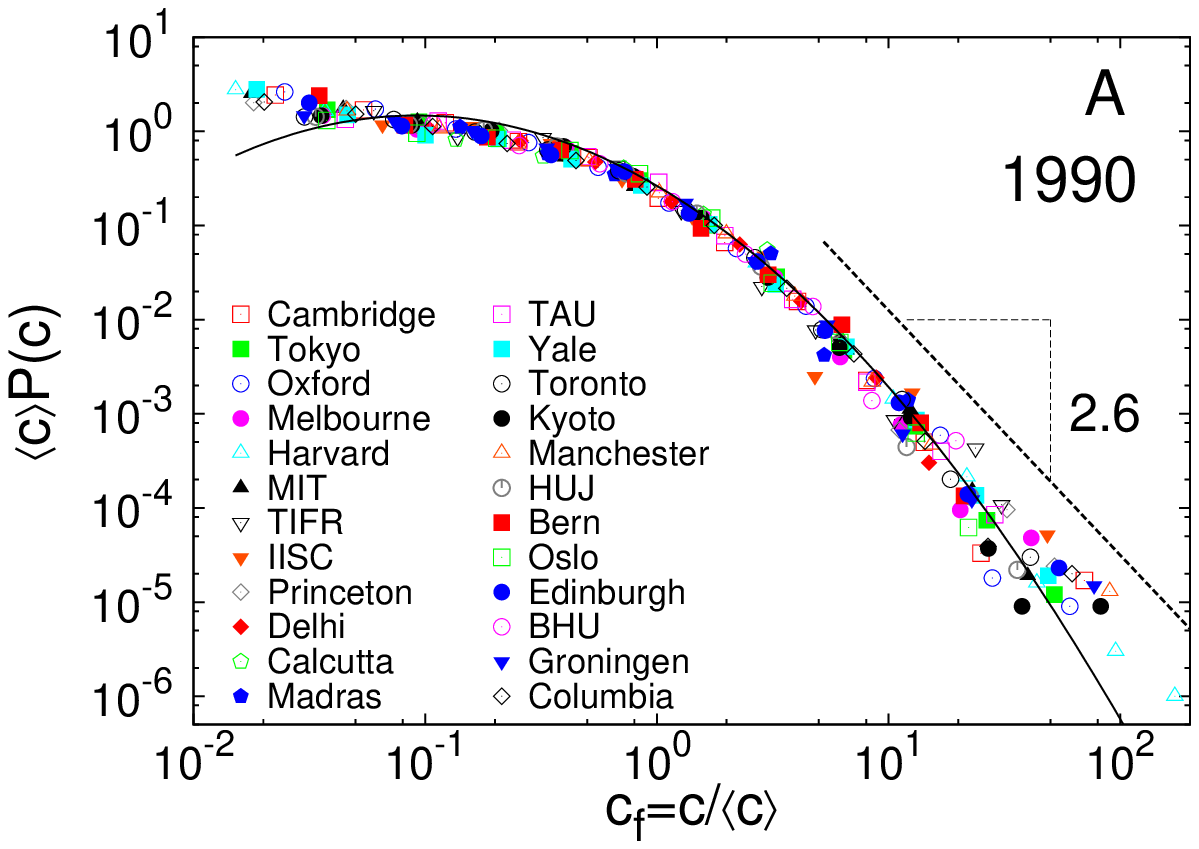}
\includegraphics[width=6.0cm]{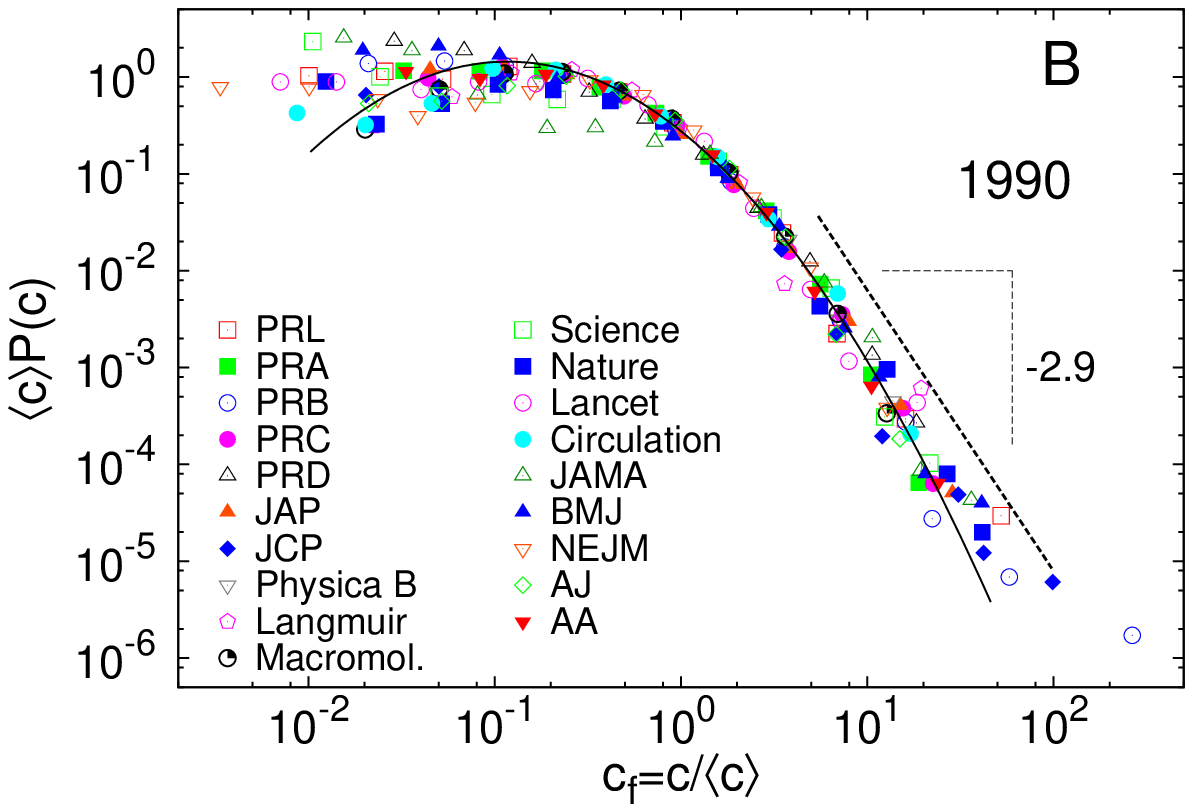}
\caption{(A) Probability distribution $P(c)$ of citations $c$ rescaled by average number 
of  citations $\langle c \rangle$
to publications from 1990 for several several academic institutions. 
The scaled distribution fits very well to a lognormal for most of its range, 
with $\mu = -0.73 \pm 0.02$, $\sigma = 1.29 \pm 0.02$.
The largest citations do not follow the lognormal behavior, and seem to follow 
a 
power law: $c^{-\alpha}$, with $\alpha = 2.8 \pm 0.2$.
(B) Probability distribution $P(c)$ of citations $c$ rescaled by average number 
of 
citations $\langle c \rangle$
to publications from 1990 for several academic journals.
The scaled distribution function fits well to a lognormal 
function with $\mu = -0.75 \pm 0.02$,  $\sigma = 1.18 \pm 0.02$, 
while $\langle c \rangle P(c) \to const.$ as $c/\langle c \rangle \to 0$
for the lower range of $c$.
The largest citations fit well to a power law: $c^{-\alpha}$, with $\alpha = 
2.9 \pm 0.3$.
Data is taken from Ref.~\cite{chatterjee2016universality}.
}
\label{fig:plosfig14}
\end{figure}

In a recent study~\cite{chatterjee2016universality}, we were able to conclude that 
the citation distributions for articles published in different journals 
(Fig.~\ref{fig:plosfig14})B, as well as from different academic institutions 
(Fig.~\ref{fig:plosfig14})A followed the same functional form, irrespective of time 
(the year they are published) and space (institution). One has to carefully scale 
the probability distributions by their average, and the rescaled curves show 
excellent scaling collapse. The most of the resulting scaling curve fits to a 
lognormal function
\begin{equation}
 F(x) = \frac{1}{x \sigma \sqrt{2\pi}} \exp \left[-\frac{(\log x - \mu)^2}{2 
\sigma^2}\right],
 \label{eq:ln}
\end{equation}
while the extreme right tail  deviates from this and seem to fit 
more to a power law with a decay exponent around $2.6-2.8$.
\begin{figure}[h]
\centering
\includegraphics[width=8.0cm]{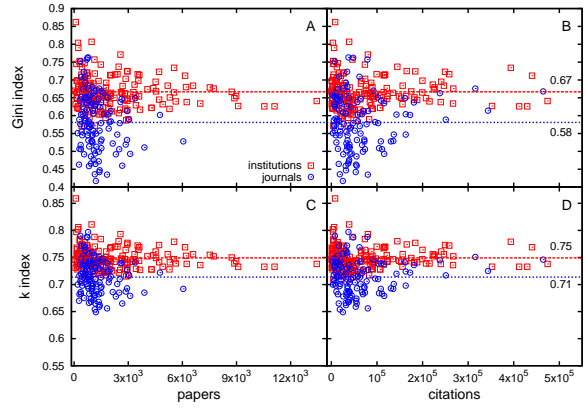}
\caption{Variation of Gini and $k$ indices  with number of papers and citations for 
academic institutions and journals.
For larger number of papers or citations, the values seem to fluctuate less or 
converge around the mean values $\bar{g}$ and $\bar{k}$ respectively.
For academic institutions, the values are $\bar{g} \approx 0.67$ for Gini and 
$\bar{k} \approx 0.75$, while for the journals, the values are $\bar{g} \approx 
0.58$ and $\bar{k} \approx 0.71$.
Figure adapted from Ref.~\cite{chatterjee2016universality}.
}
\label{fig:plosfig6}
\end{figure}
We additionally observed that for the 
academic institutions, Gini index was $g = 0.67 \pm 0.10$  and $k= 0.75 \pm 
0.04$, which means around $75\%$ citations
come from the top $25\%$ papers.
For academic journals, $g=0.58 \pm 0.15$, $k= 0.71 \pm 0.08$
which means about $71\%$ citations
come from the top $29\%$ papers.

We further noted that Gini and $k$ indices
fluctuate less around respective mean values $\bar{g}$ and $\bar{k}$ as 
the number of articles or the number of citations 
became large (Fig.~\ref{fig:plosfig6}).
For academic institutions, the values were $\bar{g} \approx 0.66$ for Gini and 
$\bar{k} \approx 0.75$. For journals, the values
are $\bar{g} \approx 0.58$ and $\bar{k} \approx 0.71$.

\section{Empirical findings on $g-k$ relationship}
The huge variety of socio-economic data suggest that there might be  
a simple relation between the two seemingly different inequality 
measures~\cite{chatterjee2017socieconomic}. 
Analysis of the following were carried out:
(i) citations of papers published from academic institutions and 
journals (data from ISI Web of Science~\cite{ISI} and reported in  
Ref.~\cite{chatterjee2016universality}), (ii) consumption expenditure data of 
India~\cite{NSSO}, Brazil~\cite{Brazil0203,Brazil0809}, Italy~\cite{Italydata},
income data from USA~\cite{IRS}, (iii) voting data from open list proportional 
elections~\cite{chatterjee2013universality} of Italy, Netherlands and Sweden,
\textit{first past the post} election data for Indian Parliamentary elections 
and Legislative Assembly elections~\cite{ECI}, 
United Kingdom~\cite{UKElec}, Canada~\cite{CAElec}, Bangladesh~\cite{BDElec}, 
Tanzania~\cite{TZElec}, and (iv) city population data from Ref.~\cite{ghosh2014zipf}.
\begin{figure*}[h]
\includegraphics[width=14.5cm]{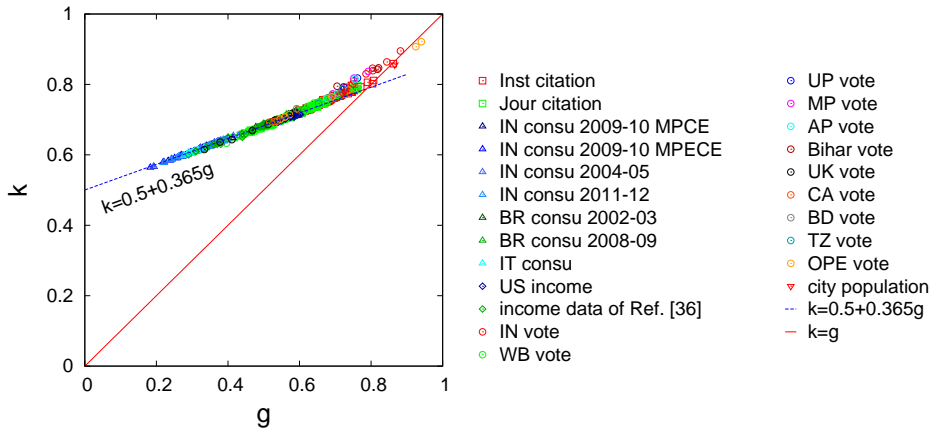}
\caption{Estimated values of $k$-index and $g$-index from the various 
datasets: citations (retrieved from ISI Web of 
Science~\cite{ISI}, analyzed in 
Ref.~\cite{chatterjee2016universality}; Inst=institutions, Jour=journals)
expenditure (data taken from 
Ref.~\cite{chatterjee2016invariant,chakrabarti2016quantifying};
IN=India, BR=Brazil, IT=Italy), income (data taken  
from Ref.~\cite{inoue2015measuring}), voting data from proportional elections 
(data taken from Ref.~\cite{chatterjee2013universality}; OPE), 
voting data from first-past-the-post elections (IN=India, WB=West Bengal, UP=Uttar Pradesh, MP=Madhya Pradesh, AP=Andhra 
Pradesh, UK=United Kingdom, CA=Canada, BD=Bangladesh, TZ=Tanzania), and city 
population (data taken from Ref.~\cite{ghosh2014zipf}).
Data details are given in Ref.~\cite{chatterjee2017socieconomic}.
The dotted straight line represents $k=0.5+0.365g$.
}
\label{fig:gk_emp}
\end{figure*}

The $g-k$ relation seems to be perfectly linear for smaller values while it 
becomes non-linear at the limit of extreme inequality, i.e., as $g$ or $k$ 
approaches unity (Fig.~\ref{fig:gk_emp}).
The most striking feature is that the data from a variety of these sources hardly 
depart from a seemingly smooth curve.

The $k$-index and $g$-index obey a linear relationship
\begin{equation}
  k = \frac{1}{2} + \gamma.g, ~~\text{for}~~ 0 \leq g \lesssim 0.70,
  \label{eq:emp}
 \end{equation}
with $\gamma =0.365 \pm 0.005$~\cite{chatterjee2017socieconomic}.

There has been a attempt to explain the slope of the $g-k$ curve for small values.
by approximating the Lorenz curve as an arc of a 
circle~\cite{chatterjee2017socieconomic}.
This linear relationship (with the value of the slope $\gamma \approx 
0.363$) can be argued to be more generally valid. If the Lorenz curve $L(x)$ in 
Fig.~\ref{fig:lorenz} is taken as a 
parabola ($L(x)=x^2$, as in case of normalized uniform distribution $P(m)$ of
income/wealth $m$; 
$L(x)=\int_0^x 2m P(m) dm$), one gets $g=2\int_0^1 (x-L(x))dx =\frac{1}{3} 
\approx 0.33$ and $1-k=L(k)=k^2$, giving $k=\frac{1}{2}(\sqrt{5} -1) \approx 
0.62$, the values of $g$ and $k$ satisfy the above relationship very well.

\section{Estimates of $g-k$ relation from kinetic exchange models}
\label{sec:model}
\begin{figure}[h]
\includegraphics[width=6.0cm]{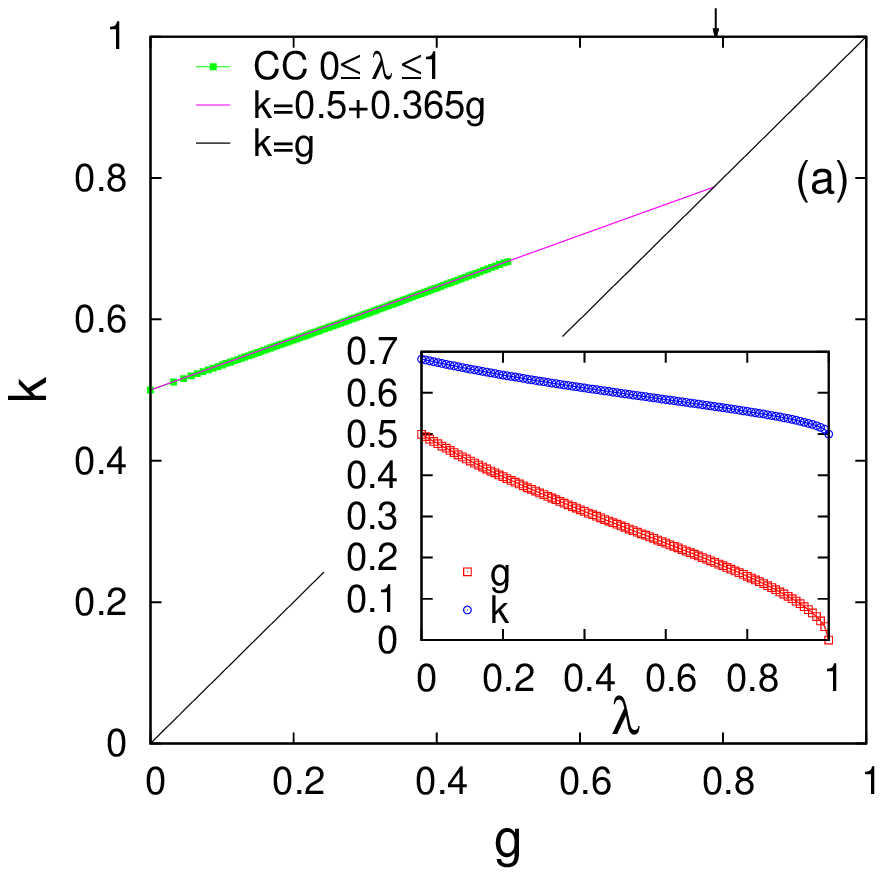}
\includegraphics[width=6.0cm]{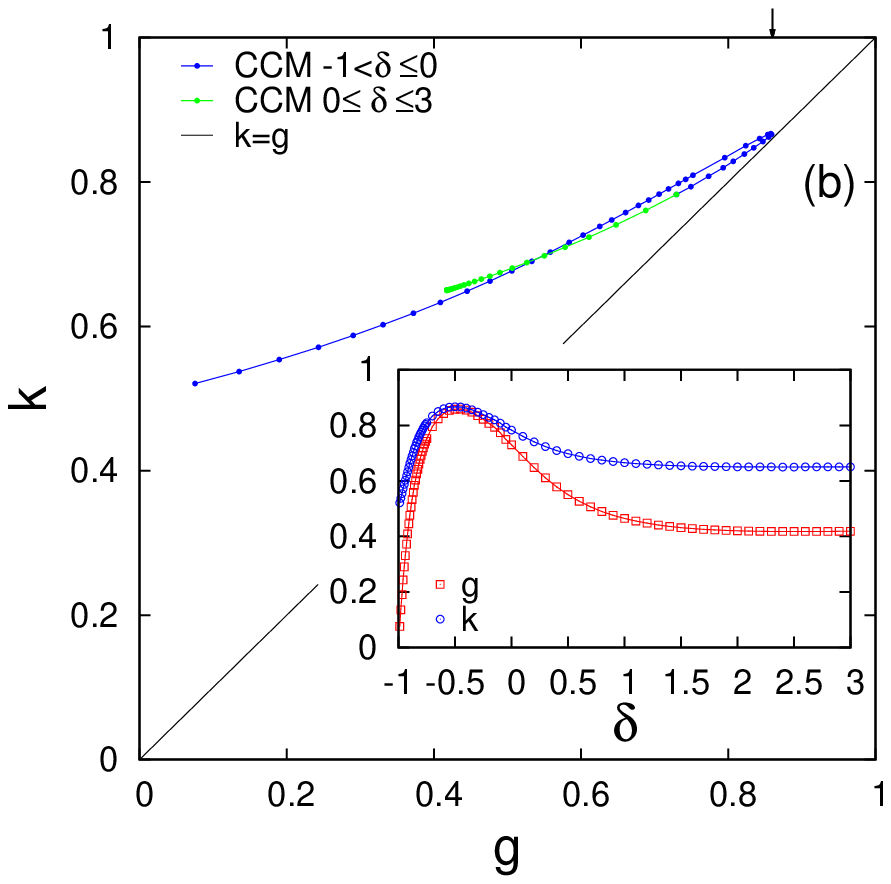}
 \caption{Monte Carlo simulation results for $g$ vs. $k$ in CC and CCM models 
(for $1000$ agents). 
 (a) For CC model, varying parameter $\lambda$.
 The inset shows the plots of $g$ and $k$ in the range of $0\le \lambda \le 1$.
 The points fit to $k= \frac{1}{2} + \gamma.g$ with $\gamma \approx 0.365 \pm 
0.005$.
 (b)  For CCM model, varying parameter $\delta$. The inset shows the 
variation of $g$ and $k$ in the range of $-1 < \delta \le 3$.
Figure adapted from Ref.~\cite{chatterjee2017socieconomic}
 }
 \label{fig:gk_models}
\end{figure}

The market models developed by physicists, specifically the kinetic exchange 
models~\cite{Chatterjee2007,chakrabarti2013econophysics} can provide an estimate of 
the relation between the inequality indices. 
In the CC model~\cite{Chatterjee2007}, where an agent retains a (same for all) 
 fraction $\lambda$ 
of their income or wealth before going for any (stochastic) exchange 
(call it trade or scattering) with another agent, the dynamics is defined by
\begin{equation}
\begin{split}
m_i(t+1) = \lambda m_i(t)+r(1-\lambda)\left[ m_i(t)+m_j(t) \right] \\
m_j(t+1) = \lambda m_j(t)+(1-r)(1-\lambda)\left[ m_i(t)+m_j(t) \right], \nonumber
\label{eq:cc}
\end{split}
\end{equation}
where $r$ is a random fraction in $[0,1]$, drawn at each time step 
(trade or exchange). 
$m_i(t)$ and $m_i(t+1)$ are the wealth of the $i$th agent at trading times $t$ 
and $(t+1)$ respectively.
The steady state distribution of 
wealth is argued to be Gamma distribution~\cite{Patriarca2004,Chatterjee2007} with 
the peak 
position 
shifting to higher income or wealth  as $\lambda$ increases ($\lambda=0$ 
corresponds to Gibbs or exponential distribution and $\lambda \to 1$ 
approaches $\delta$-function). The $g-k$ relationship for such 
distributions is found to be linear 
(Fig.~\ref{fig:gk_models}a), obeying $k= \frac{1}{2} + 
\gamma.g$ with $\gamma \approx 0.365 \pm 0.005$.

In the CCM model~\cite{chakrabarti2013econophysics,Chatterjee2007}, each agent 
$i$ has a saving fraction $\lambda$ drawn from a 
(quenched) distribution $\Pi(\lambda) = (1+\delta)(1-\lambda)^\delta$. 
Following similar stochastic dynamics as in CC model, 
\begin{equation}
\begin{split}
m_i(t+1) = \lambda_i m_i(t)+r \left[ (1-\lambda_i)m_i(t)+(1-\lambda_j)m_j(t) 
\right]\\
m_j(t+1) = \lambda_j 
m_j(t)+(1-r)\left[ (1-\lambda_i)(m_i(t)+(1-\lambda_j)m_j(t) \right], \nonumber
\label{eq:ccm}
\end{split}
\end{equation}
one gets a steady state distribution of income or wealth with power law tails 
$P(m)\sim m^{-(2+\delta)}$ for large $m$~\cite{Chatterjee2007}.
$g$ and $k$ computed for such distributions~\cite{ghosh2016inequality} are 
given in inset of Fig.~\ref{fig:gk_models}b for varying range of $\delta$. The 
$g-k$ relationship here is found to be nonlinear (see 
Fig.~\ref{fig:gk_models}b) but 
very much  around a similar linear relationship.

\section{Universality in the statistics of deaths in conflicts and disasters}
The history of human civilization has been frequently shaped by events of wars, conflicts and disasters. In recent times, the
scale of disaster events have increased remarkably. Growing population around the world has been seen as one of
the reasons for the increase in counts of people affected by disaster events.
A study on the statistics of human deaths from wars, conflicts as well
as natural disasters shows that the probability distribution of number of people killed in natural disasters
as well as man made situations exhibit similar universality in statistics with power law decay for the largest sizes,  the exponent
values being quite similar~\cite{chatterjee2016fat}, in the range of $1.5-1.8$. 
Comparing with natural disasters, where event sizes are measured in terms of physical
quantities, like the energy released in earthquake, the volume of rainfall, the land area affected in forest fires,
etc. also show striking similarities. These universal patterns in their statistics might suggest some
subtle similarities in their mechanisms and dynamics.
\begin{figure}[h]
\includegraphics[width=12.5cm]{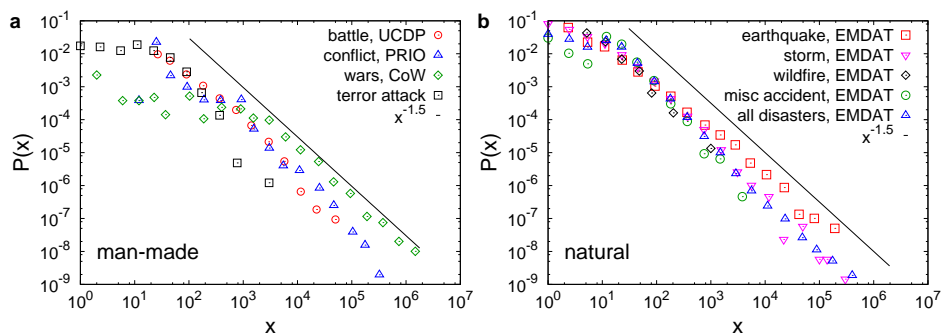}
 \caption{
The probability distributions $P(x)$ of event size $x$, measured by the number 
of human deaths in the corresponding event.
(a) \textit{Man-made events}: human deaths from conflicts 
during 1946-2008, according to the PRIO database~\cite{PRIOdata} (using lowest 
estimates),
dead according to the Correlators of Wars (CoW) 
database~\cite{CoWdata} during 1816-2007, 
dead in terror attack~\cite{Terrordatawiki} during 1910 till July 
2016, and battle deaths according to UCDP database~\cite{UCDPdata} during 
1989-2014.
Except the terrorist attack data, all these distributions seem to have a power 
law tail with similar exponents. The straight line is a guide to the exponent 
value $1.5$ for comparison. 
(b) \textit{Natural disasters}: human death from 
earthquakes, storms, wildfires, miscellaneous accidents, as well as all natural 
disasters listed in  the EMDAT database~\cite{EMDATdata} during 1900-2013.
The values of the exponents are in $1.5-1.8$ (details in Ref.~\cite{chatterjee2016fat}).
 }
 \label{fig:alldeaths}
\end{figure}

\section{Discussions on citations and relationship between inequality measures in general}
The Gini index $g$ is the most popular among economists 
and sociologists, since it gives an overall measure of the inequality in a 
society. As evident from Fig~\ref{fig:lorenz}, it requires accurate data 
for the entire Lorenz curve to provide a measure of the shaded area enclosed by it 
and the equality line. The Kolkata index $k$ 
being given by the intersection of the Lorenz curve and the cross diagonal 
to the equality line. The $g-k$ linear relationship is 
extremely robust for not so high values of inequality and fits different forms of Lorenz curve and hence,  
distributions of income, wealth, citations, etc. and this robustness is also 
observed empirically (Fig.~\ref{fig:gk_emp}).
We could even compare these findings with simple kinetic exchange models of wealth distributions,
where the scaling relation between $g$ and $k$ was found to be also true.
The $g-k$ relationship would be 
extremely useful to translate from one inequality measure to the other; since 
$1-k$ fraction of people possess precisely $k$ fraction of the total 
wealth, translation of social inequality measures into $k$-index language can  
be of major significance.

One of our recent focus had been the inequality in citations for academic institutions and journals.
Although institutions and journals have their own ranking depending on the `quality' of research and publications
that come out, get noticed and cited, we find that the form of the distribution function for citations is invariant
with respect to the average citations, holding across institutions and over time as well.
In terms of absolute inequality measures, roughly 75\% citations come from the top 25\% papers in case of academic 
institutions and 71\% citations come from the top 29\% papers for journals.

We also discussed how the inequality statistics of deaths in social conflicts or wars compare with 
those in natural disasters.

\section{Concluding remarks: Some random thoughts about prospects of institutional econophysics}

Twenty years have passed since the formal coining of the term and hence the launch 
of econophysics as a research topic (since 1995; see the entry by Barkley
Rosser on Econophysics in The New Palgrave Dictionary
of Economics~\cite{Barkleyroseerency2008}).
Furthermore, econophysics has been assigned the Physics and Astronomy
Classification Scheme (PACS) number 89.65Gh by the American
Institute of Physics. However, regular  interactions and
collaborations  between the communities of natural scientists
and social scientists are rare. Though interdisciplinary
research papers on econophysics and sociophysics are
regularly being published at a steady and healthy rate
(more than 1000 documents containing the explicit term
``econophysics" and more than 240 documents containing the
explicit term ``sociophysics" in the years 2014 and 2015
according to Google Scholar) and published mostly in physics
journals, and a number of universities (including
Universities of Leiden, Bern, Paris and London) are offering the
interdisciplinary courses on econophysics and sociophysics,
not many clearly designated professor or other faculty positions for that matter 
are available yet (except for econophysics in Universities of Leiden and
London). Neither there are any designated institutions on
these interdisciplinary fields, nor separate departments or centres of studies for 
instance. We note however, happily
in passing, a recently published highly acclaimed (``landmark" and ``masterful")
economics book~\cite{Shubik2016guidance} by Martin Shubik (Seymour Knox
Professor of  Mathematical Institutional Economics,
Emeritus, at Yale University) and Eric
Smith (Santa Fe Institute) discusses extensively on econophysics approaches
and in general on the potential of interdisciplinary
researches inspired by the developments in natural
sciences.

In view of these, it seems it is time to try for an
international centre for interdisciplinary studies on
complexity in social and natural sciences; specifically
on econophysics and sociophysics\footnote{Although the presentation in the conference (by BKC) was mainly on the materials 
discussed in the earlier sections, extensive discussions with several 
participants, including the conference organizers, had been on this point.}. The model of the
Abdus Salam International Centre for Theoretical Physics (ICTP),
Trieste (funded by UNESCO and IAEA), could surely be
helpful to guide us here.
We are contemplating if an ICTP-type interdisciplinary research
institute  could be initiated for researches on econophysics
and sociophysics.  

We note that Helbing (ETH, Zurich) and
colleagues have been trying for an European Union funded
``Complex Techno-Socio-Economic Analysis Centre'' or ``Economic
and Social Observatory" for the last five years  (see Ref.~~\cite{helbing2011white}
containing the White Papers arguing for the  proposed
centre).
We are also aware that Indian Statistical Institute had taken a decision
to initiate a similar centre in India~(see the ``Concluding Remarks'' in \cite{ghosh2013econophysics}).
Also there was an attempt for a similar Asian Centre in Singapore, initiated in Nyangyong Technological University. 
In view of some recent enthusiasms at the
Japan-India Heads of States  or Prime Minister level, and
signing of various agreements (predominantly for business
deals, infrastructure development, technical science and 
also cultural exchanges) by them, possibility of an
Indo-Japan Center for studies on Complex Systems is 
also being explored. In such bilateral (Indo-Japan) initiatives,
there are explicit Memorandum of  Understandings
already signed by the Prime Ministers. It did not have any economic or 
sociological study centres ever planned under such  bilateral
efforts.

These proposals are for regular research centres on such
interdisciplinary fields, where regular researchers will 
investigate such systems. However, in view of the extreme interdisciplinary 
nature of econophysics and sociophysics, such efforts may be complemented by 
another visiting centre model. Unlike the above-mentioned kind of 
centres therefore this proposed centre may be just a visiting centre where natural and
social scientists from different universities and institutions
of the world can meet for extended periods to discuss and
interact on various interdisciplinary issues and collaborate for such researches,
following the original ICTP model.

Here, as in ICTP, apart from a few (say, about ten to start-with)
promising young researchers on econophysics and sociophysics as
permanent faculty who will continue active research and active
visiting scientist programmes (in physics, economics and 
sociology) etc. can be pursued, The faculty members, in
consultation with the advisers from different countries,
can choose the invited visitors  and workshops or courses, on
economics and sociological complexity issues, can be
organized on a regular basis (as for basic theoretical
sciences in ICTP or in Newton Centre, Cambridge, etc.).

We think, it is an appropriate time for the healthy growth
of these ``New or Evolving Economic \& Sociological Thinkings"
including econophysics and sociophysics. We believe, Tokyo would be
the ideal location for such an International Centre.
In such new studies on social sciences, econophysics and
sociophysics in particular, Japan has already significantly
large, active and established groups and hence, Tokyo could be its
natural location.

\section*{Acknowledgment}
We are extremely thankful to Yuji Aruka and Taisei Kaizoji for sharing with us 
many enthusiastic ideas on several collaboration projects. We are also grateful to 
Sudip Mukherjee for his help in preparing the manuscript.

\end{document}